\documentstyle[prd,aps,epsf,epsfig,twocolumn]{revtex}
\begin{document}
\draft

\twocolumn[\hsize\textwidth\columnwidth\hsize\csname@twocolumnfalse\endcsname

%
\title{
Halting viruses in scale-free networks}
\author
{Zolt\'an Dezs\H{o} and Albert-L\'aszl\'o Barab\'asi}
\address{
{\it Department of Physics, University of Notre Dame, Notre Dame,
IN 46556}
}

\maketitle
\centerline{\small (Dated \today)}

\begin{abstract}
The vanishing epidemic threshold for viruses spreading on
scale-free networks indicate that traditional methods, aiming to decrease a
virus' spreading rate cannot succeed in eradicating an epidemic. We demonstrate that
policies that discriminate between the nodes, curing mostly the
highly connected nodes, can restore a finite epidemic
threshold and potentially eradicate a virus. We find that the
more biased a policy is towards the hubs, the more chance it has to bring the
epidemic threshold above the virus' spreading rate. Furthermore, such
biased policies are more cost effective, requiring
less cures to eradicate the virus.
\end{abstract}

\vspace{2pc}
]

\vspace{1cm}

\narrowtext

While most diffusion
processes of practical interest, ranging from the spread of
computer viruses to the diffusion of sexually transmitted diseases,
take place on complex networks,
the bulk of diffusion studies have focused on model
systems, such as regular lattices or
random networks\cite{nowak,havlin}.
 A series of recent results
indicate, however, that real networks significantly deviate from the structure
of these model systems\cite{reka} -- deviations that have a strong impact on the
diffusion dynamics as well.
 In particular, the
 networks responsible for the spread of computer viruses,
 such as the Internet\cite{fal} or the email network \cite{bold},
 have a scale-free topology \cite{attach}, exhibiting a power-law
degree distribution $P(k) \sim k^{-\gamma}$ where $\gamma$ ranges between $2$ and
$3$.
   Similarly, a recent study indicate that the
social network responsible for the spread of sexually transmitted
 diseases, such as AIDS, also exhibits a scale-free structure\cite{liljeros}.
 The topology of scale-free networks
 fundamentally deviate from the topology of both regular lattices and
 random networks \cite{erdos}, differences that
 impact the network's robustness and attack
 tolerance \cite{attack} or the dynamics of synchronization \cite{sink}.
 It is not unexpected, therefore, that the broad degree distribution leads
 to unexpected diffusion properties as well \cite{pastor}.

A simple model often used to study the generic features of virus spreading is
the susceptible-infected-susceptible (SIS) model.
In this model an individual is represented by a node,
which can be either "healthy" or "infected". Connections between individuals along which
the infection can spread are represented by links.
In each time step a healthy node is
infected with probability $\nu$ if it is connected to at least one infected node. At the same time
an infected node is cured with probability $\delta$, defining an effective spreading rate
$\lambda \equiv\frac{\nu}{\delta}$ for the virus.

The behavior of the SIS model is well understood if the nodes are placed on a regular
lattice or a random network \cite{nowak}. Diffusion studies indicate that viruses whose
spreading rate exceeds a critical threshold $\lambda_c$ will persist, while those under the
threshold will die out shortly. Recently, however, Pastor-Satorras and Vespignani have
shown \cite{pastor} that for scale-free networks with $\gamma\leq 3$ the epidemic threshold
vanishes, i.e. $\lambda_c=0$.
 This finding implies that on such networks even weakly
 infectious viruses can spread and prevail. This vanishing threshold
 is a consequence of the hubs -- nodes with a large
number of links encoded by the tail of power law $P(k)$.
Indeed, the hubs are in contact with a large number of nodes, and are therefore
easily infected. Once infected, they pass on the virus to a significant fraction of the
nodes in the system.

The finding that the epidemic threshold vanishes in scale-free networks has
a strong impact on our ability to control various virus outbreaks.
Indeed, most methods designed to eradicate viruses -- biological or computer based -- aim at reducing
the spreading rate of the virus, hoping that if $\lambda$ falls under
the critical threshold $\lambda_c$,
the virus will die out naturally.
With a zero threshold, while a reduced spreading rate will decrease the
virus' prevalence,
there is little guarantee that it will eradicate it.
Therefore, from a theoretical perspective viruses spreading on a scale-free network
appear unstoppable. The question is, can we take advantage of the increased
knowledge accumulated in the past few years about network topology to
understand the conditions in which one can successfully eradicate viruses?

Here we study the spreading of a virus
to which there is a cure, eradicating the virus from the node to which it is
applied to, but which does not
offer a permanent protection against the virus. If such cure is available to
all nodes, treating simultaneously all infected nodes will inevitably wipe the virus out.
However, due to economic or policy considerations the
number of available cures is often limited. This applies
to AIDS, for which relatively effective but prohibitively
expensive cures are available, unable to reach the most affected segments of population due
to economic considerations \cite{AIDS}. But it also applies to
computer viruses, where only a small fraction of users commit the
time and investment to update regularly their virus protection system.
We show that distributing the cures randomly in a scale-free
network is inneffective, being unable to alter the fundamental properties
of the threshold-free diffusion process.
However, even weakly biased strategies, that discriminating between the nodes, curing with a higher
probability the hubs than the less connected nodes, can restore the epidemic threshold.
We find that such hub-biased policies are more cost-effective as well, requiring fewer
cures than those distributing the cures indiscriminately.

{\it Curing the hubs:}
The vanishing epidemic threshold of a virus spreading in a scale-free network
is rooted in the infinite variance of
the degree distribution \cite{pastor}. Indeed, the threshold $\lambda_c$ depends on the
variance as
\begin{equation}
\lambda_c=\frac{<k>}{<k^2>}.
\end{equation}
On a regular lattice the degree distribution is a
delta function, while on a random network it follows a Poisson distribution, in both cases resulting
in a finite $<k^2>$, and therefore nonzero $\lambda_c$. In contrast, if the virus spreads on a scale-free network,
for which $P(k)$ follows a power law with $\gamma\leq 3$, the variance is infinite and the epidemic threshold is
$\lambda_c=0$.
  Therefore, to restore a finite epidemic threshold, which would allow the infection to die out,
   one needs to induce a finite
variance. As the origin of the infinite variance is in the tail of the degree distribution, dominated by
the hubs, one expects that curing {\it all} hubs with degree larger
than a given degree $k_0$ would restore a finite variance and therefore a nonzero epidemic
threshold. Indeed, if on a scale-free network nodes with
degree $k>k_0$ are always healthy, the epidemic threshold is finite and has the value\cite{unknown}
\begin{equation}
\lambda_c=\frac{<k>}{<k^2>}=\frac{k_0-m}{k_0 m} {\left(\ln\frac{k_0}{m}\right)}^{-1}.
\end{equation}

\begin{figure}[h]
\begin{center}
\epsfig{figure=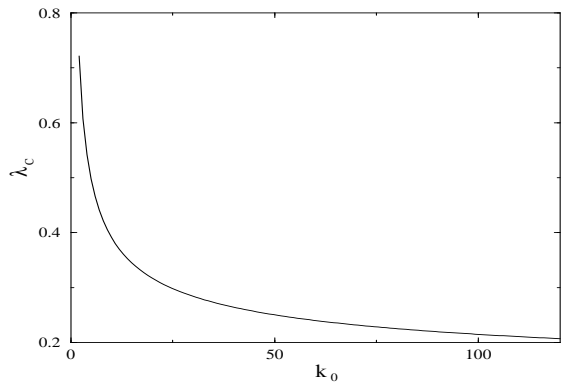,height=5cm,width=7.5cm}
\caption{The epidemic threshold as a function of $k_0$.}
\label{fig:fig1}
\end{center}
\end{figure}

This expression indicates that
the more hubs we cure (i.e. the smaller $k_0$ is), the larger the value of the
epidemic threshold (Fig.~1) \cite{imun}.
Therefore, the most effective policy against an epidemics would be to cure as many hubs as
economically viable.
The problem is that in most systems of interest
 we do not have detailed network maps, thus we cannot
effectively identify the hubs. Indeed, we do not know the number of sexual partners for
each individual in the society, thus we cannot identify the social hubs that should be
cured if infected. Similarly, on the email network we do not know which
email accounts serve as hubs, as these are the ones that, for the benefit of all email users,
should always carry the latest anti-virus
software.

Short of a detailed network map, no method aiming to identify and cure
 the hubs is expected to succeed at its goal of finding {\it all} hubs with degree larger than a given $k_0$.
Yet, policies designed to eradicate viruses could attempt to identify and cure as many hubs as possible.
Such biased policy will inevitably be inherently imperfect,
as it might miss some hubs, and falsely identify some smaller nodes
as hubs. The question is, however, would a policy
biased towards curing the hubs, without a guarantee that it can identify
all of them, succeed at restoring the epidemic threshold?

To investigate the effect of incomplete information about the hubs we assume that the likelihood of
identifying and administering
a cure to an infected node
with $k$ links in a given time
frame depends on the node's degree as $k^{\alpha}$, where $\alpha$ characterizes the policy's ability to identify hubs.
In this framework $\alpha=0$ corresponds to random cure distribution,
which is expected to have zero epidemic threshold while $\alpha=\infty$
corresponds to an optimal policy that treats all hubs with degree larger than $k_0$.
 Within the framework of the SIS model we assume that each node is infected
 with probability $\nu$, but each infected node is cured with
 probability $\delta=\delta_0 k^{\alpha}$, becoming again susceptible
 to the disease. We define the spreading rate as $\lambda = \frac{\nu}{\delta_0}$.
As each healthy node is susceptible again to the disease,
a node can get multiple cures during a simulation.

\begin{figure}[h]
\begin{center}
\epsfig{figure=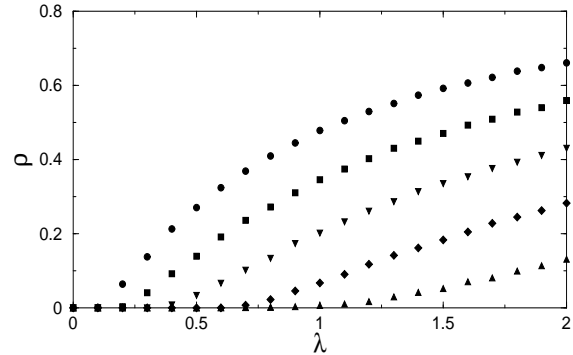,height=5cm,width=7.5cm}
\caption{          Prevalence, $\rho$, measured as the fraction of infected nodes in function
           of the effective spreading rate $\lambda$ for
          $\alpha=0$({\large o}), 0.25($\Box$), 0.50($\nabla$), 0.75($\Diamond$) and 1($\triangle$),
 as predicted by Monte-Carlo simulations using the SIS model on a scale-free
       network with N=10,000 nodes.
}
\label{fig:fig2}
\end{center}
\end{figure}

 We place the nodes on a scale-free network \cite{pref}
 and initially infect half of them. After a transient regime the system
 reaches a steady state, characterized by a constant average density of infected nodes, $\rho$,
 which depends on both the spreading rate $\lambda$ and $\alpha$ (Fig.~2).
 The $\alpha=0$ limit corresponds to random immunization in which case the
epidemic threshold is zero.
As treating only the hubs will restore
the nonzero epidemic threshold, for $\alpha=\infty$ we expect a nonzero $\lambda_c$.
 Yet, the numerical simulations indicate that we have a finite $\lambda_c$
well before the $\alpha=\infty$ limit. Indeed, as Fig.~2 shows, $\lambda_c$ is
clearly finite for $\alpha=1$ and so is for smaller value of $\alpha$ as well. The numerical simulations
do not give an unambiguous answer the crucial question: Is there a critical value of $\alpha$ at which a
finite $\lambda_c$ appears,
or for any any nonzero $\alpha$ we have a finite $\lambda_c$?

{\it Mean-field theory:}
To interpret the results of the numerical simulations we studied the effect of a biased
policy using the mean-field continuum approach \cite{nowak,pastor}.
Denoting by $\rho_k(t)$ the density of infected nodes with connectivity $k$, the
time evolution of $\rho_k(t)$ can be written as \cite{pastor}
\begin{equation}
 \partial_t\rho_k(t)=-\delta_0k^{\alpha}\rho_k(t)+\nu(1-\rho_k(t))k\theta(\lambda).
\end{equation}
The first term in the r.h.s. describes the probability that an infected node is cured,
 and it is therefore proportional to the
number of infected nodes $\rho_k(t)$ and the probability $\delta_0k^{\alpha}$ that a node
with $k$ links will be selected for a cure.
The second term is the probability that a healthy node with $k$ links is infected,
proportional to the infection rate ($\nu$), the number of
links ($k$), the number of healthy nodes with $k$ links ($1-\rho_k(t)$),
and the probability $\theta(\lambda)$ that a given link points to an infected node.
The probability $\theta(\lambda)$ is proportional to $kP(k)$,
 therefore it can be written as
\begin{equation}
\theta(\lambda)=\sum_k \frac{kP(k)} {\sum_s{sP(s)}}\rho_k.
\end{equation}
Using $\lambda = \frac{\nu}{\delta_0}$ and
imposing  the $\partial_t\rho_k(t)=0$ stationary condition we find the stationary density as
\begin{equation}
\rho_k=\frac{\lambda\theta(\lambda)}{k^{\alpha-1}+\lambda\theta(\lambda)}.
\end{equation}
Combining Eq.(4) and Eq.(5) and using the fact that the connectivity distribution $P(k)=2m^2/k^{-3}$ for the scale-free
 network\cite{attach},
we obtain:
\begin{equation}
m\lambda \int_m^{\infty}\frac{dk}{k^2(k^{\alpha-1}+\lambda\theta(\lambda))}=1.
\end{equation}
The average density of infected nodes is given by
\begin{equation}
\rho(\lambda)=\sum_{k}P(k)\rho(k)=2m^2\lambda\theta(\lambda)\int_m^{\infty}\frac{dk}{k^3(k^{\alpha-1}+\lambda\theta(\lambda))}
\end{equation}
Equations (6) and (7) allow us to calculate the average density of infected nodes for any value of $\alpha$.
For $\alpha=0$ they reduce to the case studied in Ref.\cite{pastor} giving $\lambda_c=0$.
For $\alpha=1$ we can solve (6), and using (7) we obtain
\begin{equation}
\rho(\lambda)|_{\alpha=1}=\frac{\lambda-1}{\lambda},
\end{equation}
which indicates
that for $\alpha=1$ the epidemic threshold is finite, having the value $\lambda_c(\alpha=1)=1$ \cite{imun}.
\begin{figure}[h]
\begin{center}
\epsfig{figure=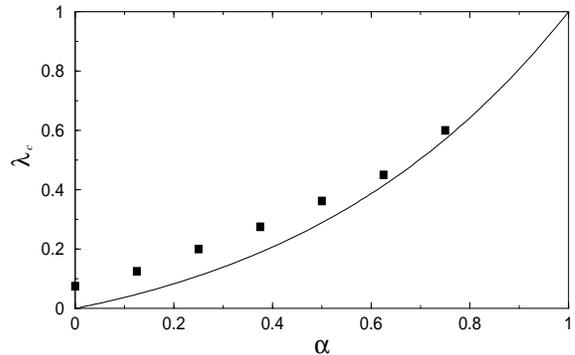,height=5cm,width=7.5cm}
\caption{           The dependence of the epidemic threshold $\lambda_c$ on $\alpha$ as predicted by
 our calculations (continuous line) based on the continuum approach, and by the numerical
simulations based on the SIS model (boxes). The small deviation between the numerical results
and the analytical prediction is due to the uncertainty in determining the precise value of the
threshold in Monte-Carlo simulations.
}
\label{fig:fig3}
\end{center}
\end{figure}
To determine the epidemic threshold as a function of $\alpha$ we need to solve the $\rho(\lambda)=0$
equation. While we cannot get $\rho(\lambda)$ for arbitrary values of $\alpha$, we
can solve Eq.(6) in $\lambda$ using that at the threshold $\lambda=\lambda_c$  we have $\theta(\lambda_c)=0$.
In this case (6) predicts that the epidemic threshold depends on $\alpha$ as
\begin{equation}
\lambda_c=\alpha m^{\alpha-1}.
\end{equation}
For $\alpha=0$ we recover $\lambda_c=0$, confirming that random immunization cannot eradicate an
 infectious disease.
 For $\alpha=1$ Eq.(9) predicts that the epidemic threshold is $\lambda_c=1$, in agreement with (8).
 Most important, however, Eq.(9) indicates
 that $\lambda_c$ is nonzero for any positive $\alpha$, i.e., any policy that is biased towards curing the hubs
 can restore
 a finite epidemic threshold. Furthermore, policies with larger
 $\alpha$ are expected to be more likely to lead to the eradication of the virus,
  as they result in larger $\lambda_c$ values.
Therefore, Eq.(9) indicates that a potential avenue to eradicating a virus is to increase
the effectiveness of identifying and curing the hubs. Indeed, if the virus has a fixed
spreading rate, increasing $\alpha$ could increase $\lambda_c$ beyond $\lambda$,
thus making possible for the virus to die out naturally.
To test the validity of prediction (9) we determined numerically the
$\lambda(\alpha)$ curve from the simulations shown in Fig.~2.
As Figure 3 shows, we find excellent agreement between the simulations and the
analytical prediction (9).

{\it Cost-effectiveness:}
A major criteria for any policy designed to combat an epidemic is its cost-effectiveness.
Supplying cures to all nodes infected by a virus is often prohibitively expensive.
Therefore, policies that obtain the largest effect with the smallest number of administered cures are more desirable.
 To addres the cost-effectiveness of a policy targeting the hubs we calculated
the number of cures administered in a time step per node for different
values of $\alpha$.
Figure 4 indicates that
increasing the policy's bias towards the hubs by allowing a higher value for $\alpha$ decreases
rapidly the
 number of necessarily cures. Therefore, policies that distribute the cures mainly to
  the nodes with more links are more cost effective than those that spread the cures randomly, blind
  to the node's connectivity.
We can understand the origin of the rapid decay in $c(\alpha)$ by noticing that
the number of cures administered per unit
time is proportional to the density of infected nodes.
From Fig.~2 we see that for a given value of
the spreading rate the prevalence is decreasing as $\alpha$ increases,
 decreasing the number of necessarily  cures as well.
\begin{figure}[h]
\begin{center}
\epsfig{figure=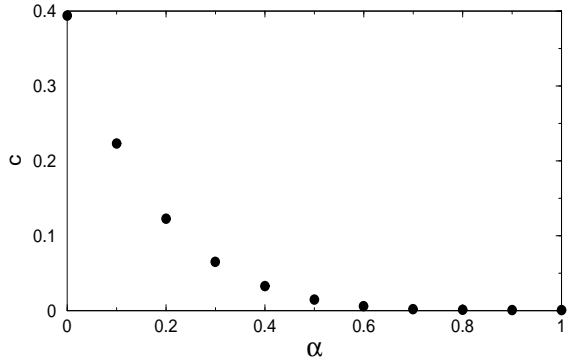,height=5cm,width=7.5cm}
\caption{  The number of cures, c, administered in an unit time per node
 for different values of $\alpha$. The rapidly decaying
$c$ indicates that more successful is a policy in selecting and curing hubs (larger is $\alpha$),
fewer cures are required for a fixed spreading rate ($\lambda=0.75$).
For $\alpha=0$ the number of cures is calculated by $c=\nu/(\nu+\delta)=
\lambda/(1+\lambda)$ which gives $c=0.43$, which value is
in good agreement with the numerical results.
}
\label{fig:fig4}
\end{center}
\end{figure}
In summary, our numerical and analytical results indicate that
targeting the more connected infected nodes
can restore the epidemic threshold, therefore making possible the
eradication of a virus. Most important, however,  is the finding that even moderately
successful policies
with small $\alpha$ can lead to a nonzero epidemic threshold.
As the magnitude of $\lambda_c$ rapidly decreases with $\alpha$,
the more effective a policy is at identifying and curing the hubs of a scale-free network, the higher are
its chances of eradicating the virus.
Finally, the simulations shows that a biased treatment policy is not only more efficient
but it is also less expensive than random immunization. These results, beyond
improving our understanding of the basic mechanisms of virus spreading, could also
offer important input into designing effective policies to eradicate computer or biological
infections.



\begin{references}
\bibitem {pref} To generate the scale-free network,
we start from a small number of
nodes ($m_0$) and each time step we add a new node to the network, with $m$ links that are connected to
 an old node
$i$ with $k_i$ links according to the probability $k_i/\sum_j kj$. After iterating the system we obtain a network
with connectivity distribution $P(k)\sim k^{-3}$ and average connectivity $<k>=2m$\cite{attach}.
\end{references}
\end{document}